# Reaction dynamics for the $Cl(^2P) + XCl \rightarrow XCl + Cl(^2P)$ (X = H ,D ,Mu) reaction on a high-fidelity ground state potential energy surface

Qiang Li[a,1)], Mingjuan Yang[c,d,1)], Hongwei Song[c,*], Yongle Li[a,b,*]

*a* Department of Physics, International Centre for Quantum and Molecular Structures, Shanghai University, Shanghai 200444, China

*b* Zhejiang Laboratory, Hangzhou 311100, China

*c* State Key Laboratory of Magnetic Resonance and Atomic and Molecular Physics, Innovation Academy for Precision Measurement Science and Technology, Chinese Academy of Sciences, Wuhan 430071, China

*d* University of Chinese Academy of Sciences, Beijing 100049, China

1) equally contributed.

## Abstract

In this work, the dynamics of a prototypical heavy-light-heavy abstract reaction, $Cl(^2P) + HCl \rightarrow HCl + Cl(^2P)$, is investigated by both constructing new potential energy surface (PES) and rate coefficient calculations. Both permutation invariant polynomial neural network (PIP-NN) method and embedded atom neural network (EANN) method, based on ab initio MRCI-F12+Q/AVTZ level points, are used for obtaining globally accurate full-dimensional ground state PES, with the corresponding total root mean square error (RMSE) being only 0.043 and 0.056 kcal/mol, respectively. And this is also the first application of EANN in gas phase bimolecular reaction. The saddle point of this reaction system is confirmed to be nonlinear. Comparison with both the energetics and rate coefficients obtained on both PESs, we find the EANN is reliable in dynamic calculations. A full-dimensional approximate quantum mechanical method, ring-polymer molecular dynamics (RPMD) with Cayley propagator, is employed to obtain the thermal rate coefficients and kinetic isotopic effects of title reactions $Cl(^2P) + XCl \rightarrow XCl + Cl(^2P)$ (H, D, Mu) on both new PESs, and the kinetic isotope effect (KIE) is also obtained. The rate coefficients reproduce the experimental results at high temperatures perfectly, but with moderate accuracy at lower temperatures, but the KIE is with high accuracy. The similar kinetic behavior is supported by quantum dynamics using wave packet calculations as well.

# I. Introduction

The Cl($^2$P) + HCl→ClH + Cl($^2$P) reaction is a typical hydrogen transfer reaction and an important prototype of so-called heavy-light-heavy reaction due to its extremely small skewing angle (13.6°),[1] which may be expected to show many interesting dynamical phenomena, such as tunneling and recrossing.[2] Therefore, the kinetics of it and its isotopic analogs, Cl($^2$P) + XCl, (X=H, D, Mu) have been under extensive theoretical studies for decades.[3-8] Correspondingly, there are several different potential energy surfaces (PESs) are released. Such as, there are three London-Eyring-Polanyi-Sato (LEPS) type PESs for the ground state, with different Sato parameters, by Bondi *et al.* (named as BCMR),[3] Avigdor Persky *et al.*[4] and Schatz *et al.*(named as PK3 PES),[5] all the above three are with a collinear transition state (TS), and potential barrier as 8.5 kcal/mol. Another one is the diatomics-in-molecules + 3 Center (DIM-3C) PES by I. Last and M. Baer.[6] The TS on DIM-3C PES is also collinear, and with potential barrier as 8.3 kcal/mol. It's found that BCMR PES gave highly rotational excitation products, by studies of quasi-classical trajectory (QCT) and quantum scattering dynamics.[7] Later there are a pair of PESs based on ab initio calculations are released. The first one is from D. Truhlar's group.[1] was fitted by rotated-Morse-oscillator-spline (RMOS) based on ~5500 *ab initio* points using the polarization configuration-interaction (POL-CI) method.[9] Although the TS from POL-CI calculation was not collinear, it was set to be collinear on the PES, with a potential barrier as 10.4 kcal/mol. The dynamics of the title reaction, including thermal rate coefficient and kinetic isotope effect (KIE), were also calculated on it in the temperature range of 200 K-2400 K by the original canonical variational theory (CVT) and the improved canonical variational theory (ICVT), after scaling the potential energy by a factor 1.42, the calculated rate coefficients were in good agreement with the experimental values. Another ab initio PES is constructed by Dobbyn *et al.* (named as DCBKS PES),[8] at the level of the theory of restricted open shell coupled cluster singles doubles with perturbation triples (RCCSD-T)[10] and multireference configuration interactions (MRCI),[11, 12] then fitted by the rotated-Morse cubic-spline function. The DCBKS PES contains three electronic states, and its TS is nonlinear. This nonlinear feature was attributed as the repulsion of both p-polarized orbitals on Cl atoms. But the potential barrier of TS was also scaled by a factor of 0.815 to match the calculated rate coefficients from QCT to the experimental values.[13] Recently, our group[14] calculated the thermal rate coefficients and KIE using an approximate full-dimensional quantum mechanical method named ring polymer molecular dynamics (RPMD) method on LEPS PES[15, 16]. The results of RPMD are also consistent with those of other theoretical approaches such as ICVT and quantum dynamics (QD). For $Cl(^2P)$ + DCl reactions, the RPMD rate coefficients at higher temperatures are very accurate compared with the experimental results. However at 312.5 K the RPMD results are slightly lower than the experimental values, although it still close to all values from other theoretical methods. This may stem from the inaccuracy of the LEPS PES, since during our previous experience, the quality of PES used is essential for RPMD calculations. And since both the geometry and energetics of TS are not well defined from above discussion, it's needed to prepare accurate PES from high-level quantum chemical method.

To unveil the dynamics of the title reaction, it is usually essential to build an accurate global PES, which can be achieved by fitting a large number of high-level ab initio energy points. According to the previous work of G. Schatz[8], the MRCI is necessary to calculate the energy of sampled configurations, due to the multiple electronic states features of the reaction system. And for fitting the PES, recently there is a novel method named embedded atom neural network (EANN) proposed by Bin Jiang's group[17]. It is extendable for high dimensional bimolecular reactions when with active learning technique. So, we choose the EANN in this work. To test the performance of the EANN method on constructing potential energy surfaces of bimolecular reactions, we have also adopted the standard permutation invariant polynomial-neural network (PIP-NN) method[18-20], which has already been demonstrated to be suitable in fitting polyatomic bimolecular reactions and widely used, such as OH + $H_2O$,[21] H/Cl/OH + $CH_4$,[22-24] $N_2O$ + $C_2H_2$,[25] OH + SO[26]. As result, two new PESs are developed, one is based on PIP-NN with 5986 points, the other is based on EANN with 6515 points. The fitting error is 0.043 kcal/mol for PIP-NN PES, and is 0.056 kcal/mol for EANN, separately. The detailed analysis shows that the geometries, harmonic frequency and energy of stationary point, can be accurately reproduced by both PIP-NN PES and EANN PES. The Cayley-RPMD calculation is then performed to reveal the kinetics of title reaction and to compare it with other theoretical and experimental values, including our previous RPMD work based on the empirical LEPS[14]. Our results in this work are closer to the experimental values, and the results for isotope H is also validated by quantum dynamics using wave packets. This work is organized as follows. The PESs employed in the current work and the related theories and calculation details are introduced in section II. The results are presented and discussed in Section III. The final summary is contained in Section IV.

# II. Theory

## II.A PIP-NN PES

All electronic calculations in this work were performed using MOLPRO2015.[27] The geometries, energies and harmonic frequencies of all stationary points of the reactants, products and TSs were obtained at the MRCI-F12+Q levels.[28-30] The method is proved to give reliable potential energy surfaces since the higher energy accuracy.

The initial data set is initially sampled at three bond lengths $R_{\mathrm{HCl_1}}$, $R_{\mathrm{HCl_2}}$, and $R_{\mathrm{Cl_1Cl_2}}$ in the range of 0.8-20 Å. Further improvement of the potential energy surface by adding points to the area around the stationary points and to the reaction path. Therefore, we obtain a primitive potential energy surface.

Based on this primitive potential energy surface, RPMD calculates from 300-1500 K for different temperatures. Exploration of dynamically relevant regions verifies the performance of the potential energy surface, which behaves unreliably in regions with lacking points. Therefore, data points in these regions are sampled to repair the potential energy surface. This process is repeated over and over again to improve the potential energy surface until all relevant dynamical results converge. To improve sampling efficiency, only those points that are not close to the existing dataset are added, using the generalized Euclidean distance $\chi(\{r_i\},\ \{r_i^{'}\}) = \sqrt{(\sum_i^3 \left|\vec{r_i} - \vec{r_i^{'}}\right|^2)} < 0.08$ Å which is used to determine the interatomic distance between a data point $\{\vec{r_i^{'}}\}$ in an existing data set and a new data point $\{\vec{r_i}\}$ .This potential energy surface is examined by examining the properties of the stationary point, such as geometries, energies, harmonic frequencies, and minimum energy paths. This potential energy surface gradually improves with the addition of new points. After a few iterations, the result of this potential energy surface is converged.

a total of 5986 points were calculated at the MRCI-F12+Q/AVTZ level and fitted by the permutation invariant polynomial-neural network method(PIP-NN) with two hidden layers,

$$V = b_1^3 + \sum_{k=1}^{K}\left(\omega_{1,k}^{(3)} \cdot f_2\left(b_k^{(2)} + \sum_{j=1}^{J}\left(\omega_{k,j}^{(2)} \cdot f_1\left(b_j^{(1)} + \sum_{i=1}^{I}\omega_{j,i}^{(1)} \cdot \mathrm{G}_i\right)\right)\right)\right) \quad (1)$$

Where $I$ indicates the number of PIPs as the input layer, $j$ and $K$ are the number of neurons in the two hidden layers, respectively; $f_i(i=1,2)$ are the transfer function of the two hidden layers; $\omega_{j,i}^{l}$ is the weight connecting the $i$th neuron of ($l$-1)th layer and the $j$th neuron of the $l$th layer; $b_j^{(l)}$ are the biases of the $j$th neuron of the $l$th layer; The fitting parameters $\omega_{j,i}^{l}$ and $b_j^{(l)}$ are iterated continuously by non-linear least squares fitting of NN using the root mean square error (RMSE) as the performance function: $\mathrm{RMSE} = \sqrt{\sum_{i=1}^{N_{data}}(E_{fit}^{i} - E_{target}^{i})^2 / N_{data}}$. The input layer of NN consists of low-order PIPs, namely, symmetrized monomials of Morse-like variables of internuclear distances, $G = \hat{S}\prod_{i<j}^{3} p_{i,j}^{l_{ij}}, p_{ij} = \exp(-r_{ij}/\alpha)(\alpha = 1\ \text{Å})$ ,and $\hat{S}$ the symmetry operator which contains all the permutation operations between two identical chlorine atoms in the system.[31] In this work, all PIPs up to the maximum order of 3, resulting in 13 terms ($I$ = 13), are used as the input layer of NN.

## II.B Embedded Atom Neural Network Potentials

Although the PIPNN method works well for constructing potential energy surfaces, it is difficult to extend to the large system containing many atoms, which is caused by too many PIPs. Thus, Using the embedded atom neural network (EANN) method,[17] one can construct a surface of high-dimensional potential energy in which the total energy is calculated based on atomic energies. Specifically, atoms are embedded in an environment of other atoms, and their atomic energy is derived from an atomic neural network based on nonlinear transformations of electron densities,

$$E = \sum_{i=1}^{N} E_i = \sum_{i=1}^{N} NN_i(\boldsymbol{\rho}^i) \tag{2}$$

In Eq. 2, $\boldsymbol{\rho}_i$ is a density-like structural descriptor that may be constructed simply by atomic orbitals of Gaussian type centered around neighboring atoms,

$$\varphi_{l_x l_y l_x}^{\alpha, r_s}(\mathbf{r}_{ij}) = x^{l_x} y^{l_y} z^{l_z} \exp(-\alpha |r - r_s|^2) \tag{3}$$

where $\mathbf{r}_{ij}$ is the Cartesian coordinate vector of embedded atom $i$ with respect to neighbouring atom $j$, and r is its norm; $l_x$, $l_y$, and $l_z$ are the projections of angular momentum along the x, y, and z axes, respectively, $L = l_x + l_y + l_z$ is the total angular momentum; $\alpha$ and $r_s$ are parameters that control the radial distribution of the atomic orbital. each component of $\boldsymbol{\rho}^i$ is determined by a linear combination of the addition of atomic orbitals on nearby atoms,

$$\rho_{L,\alpha,r_s}^i = \sum_{l_x l_y l_z}^{l_x + l_y + l_z = L} \frac{L!}{l_x! l_y! l_z!} \left( \sum_{j=1}^{n_{\text{atom}}} c_j \varphi_{l_x l_y l_z}^{\alpha, r_s}(\mathbf{r}_{i,j}) f_c(\mathbf{r}_{ij}) \right) \tag{4}$$

Where $n_{atom}$ is the number of atoms around the embedded atom with a specified cutoff radius $(r_c)$ and cutoff function $f_c(\mathbf{r}_{ij})$ that decays smoothly to zero at $r_c$. And $c_j$ is the element dependent coefficient, which is optimized during training. Moreover, the embedded atom neural network (EANN) method is more efficient thanks to the linear scaling of the density-like descriptor with respect to the number of atoms around the center of each atom. This embedded atom neural network (EANN) approach has been successfully applied to molecular and periodic systems,[32-35] including several typical gas-surface reactions.[36, 37]

## II.C RPMD reaction rate theory

All calculations are performed using the RPMD rate theory implemented in the RPMDrate code[38]. Since there have been lots of review articles about it,[39, 40] here we only give a brief summary related closely to the current work. For the title reaction, this Hamiltonian can be written as follows:

$$\hat{H} = \sum_{i=1}^{3} \frac{|\hat{\boldsymbol{p}}_i|^2}{2m_i} + V(\hat{\boldsymbol{q}}_1, \hat{\boldsymbol{q}}_2, \hat{\boldsymbol{q}}_3) \tag{5}$$

Where $\hat{\boldsymbol{p}}_i$, $\hat{\boldsymbol{q}}_i$ and $m_i$ are the momentum operator, the position operator and the mass of the $i$th atom, respectively. Taking advantage of classical isomorphism between quantum systems and the ring polymer, each quantum particle is represented by a necklace formed by $n$ classical beads connected by a harmonic potential:[15, 41]

$$H(\boldsymbol{p},\boldsymbol{q}) = \sum_{i=1}^{3}\sum_{j=1}^{n}\left(\frac{|\boldsymbol{p}_i^{(j)}|^2}{2m_i} + \frac{1}{2}m_i\omega_n^2|\boldsymbol{q}_i^{(j)} - \boldsymbol{q}_i^{(j-1)}|^2\right) + \sum_{j=1}^{n}V(\boldsymbol{q}_1^{(j)},\boldsymbol{q}_2^{(j)},\boldsymbol{q}_3^{(j)}) \tag{6}$$

Where for each atom, $\boldsymbol{q}_i^{(0)} = \boldsymbol{q}_i^{(n)}$,and the force constant between adjacent beads is given by $\omega_n = (\beta_n\hbar)^{-1}$ with the reciprocal temperature of the system $\beta_n = (nk_BT)^{-1}$.

Then the Bennett-Chandler factorization[42, 43] is used to calculate the quantum transition state theory (QTST) part and transmission coefficient based on such Hamiltonian:[38-40]

$$k_{\text{RPMD}} = k_{\text{QTST}}(T;\xi^{\neq})\kappa(t\to\infty;\xi^{\neq}) \tag{7}$$

This first factor of the above equation represents the static contribution while this second factor is the dynamic correction.

Here $k_{\text{QSTS}}(T;\xi^{\neq})$ is the centroid-density QTST rate coefficient,[16, 44] evaluated at the maximum of the free energy barrier, $\xi^{\neq}$, along the reaction coordinate $\xi(\boldsymbol{q})$. In practice, it is calculated from the centroid potential of the mean force (PMF):[38-40]

$$k_{\text{QTST}}(T,\xi^{\neq}) = 4\pi R_\infty^2(2\pi\beta\mu_{\text{R}})^{-1/2}e^{-\beta[W(\xi^{\neq})-W(0)]} \tag{8}$$

Where $\mu_{\text{R}}$ is the reduced mass between the two reactants, $\beta=(k_{\text{B}}T)^{-1}$ is the reversed absolute temperature multiply the Boltzmann factor, and $W(\xi^{\neq})-W(0)$ is the free-energy difference which is obtained via umbrella integration along the reaction coordinate.[38, 45]

The second factor $\kappa(t\to\infty;\xi^{\neq})$ is named the transmission coefficient, which provides dynamical correction and is calculated by the ratio between long-time limit and zero-time limit of the flux-side correlation function:

$$\kappa(t\to\infty;\xi^{\neq}) = \frac{c_{fs}^{(n)}(t\to\infty;\xi^{\neq})}{c_{fs}^{(n)}(t\to 0_+;\xi^{\neq})} \tag{9}$$

Which captures the recrossing of the TS region and ensures that the obtained RPMD rate coefficient results do not depend on the choice of the dividing surface.[16]

It should be noted that the final RPMD rate coefficients are corrected by an electronic partition function ratio of the following form:

$$f(T)=\frac{Q_{\text{elec}}^{\text{TS}}}{Q_{\text{elec}}^{\text{reactants}}}=\frac{1}{2+\exp(-\beta\Delta E)} \tag{10}$$

to account for the spin-orbit splitting of $\mathrm{Cl}(^{2}\mathrm{P}_{1/2,3/2})(\Delta E=882\ \mathrm{cm}^{-1})$ [5, 14, 46].

In addition, when only one bead is used, the results from RPMD will reduce to the classical limit. In this limit, the static and dynamic components become the same as the classical transition state theory (TST) rate coefficient and the classical transmission coefficient, respectively. Therefore, these quantities determine the limits at which quantum effects, such as ZPE and tunneling effects, can be evaluated by using more beads. The minimum number of beads considering the quantum effect can be given by the following formula:[47]

$$n_{\min}=\beta\hbar\omega_{\max} \tag{11}$$

Where $\omega_{\max}$ is the largest vibration frequency in the system. In this work, the convergence is tested with increasing number of beads, and the numbers that yield converged results of PMF are chosen at different temperatures. In the Supporting Information, Figure S2 shows the convergence of PMF curves at 312.5 K obtained from different numbers of beads.

Additionally, there is a critical temperature named cross-over temperature[48], $T_c$, which also needs to be considered:

$$T_c=\hbar\omega_b/(2\pi k_B) \tag{12}$$

Where $i\omega_b$ is the imaginary frequency of the reaction system in the TS. The reaction system temperature is lower than $T_c$, which is considered as deep-tunneling region, the error of RPMD results would become large. Enough beads are needed to obtain accurate results. The cross-over temperature for the title reaction is $T_c$=348 K.

For obtaining PMF from RPMD, the umbrella integral[45, 49] method is employed, and the reaction coordinates are divided into a series of windows. For all reactions, the range of the reaction coordinates is set to $\xi\in$[-0.05,1.05], the interval between adjacent windows is set to 0.01, so the total number of windows is 110. In addition, the force constant for the three reactions is set to 0.1 ($T$/K) eV. At each sampling window, this system is first equilibrated by 2 ps, followed by a production run (1.2 ns split into 60 sampling trajectories). This Anderson thermostat[50] was used in all the simulations, the details are in the supporting information. The ring-polymer equations of motion were integrated in Cartesian coordinates using a highly efficient Cayley propagator[51, 52] with a time step of 0.5 fs. The convergence of the choice of time step is also tested, as shown in the Figure S3 and Table S2 of Supporting Information.

After the position of the free energy barrier is determined from the PMF calculation at each temperature, the transmission coefficient is calculated at this position. This was initialized by running a long (60 ns) parent trajectory, using the SHAKE algorithm to fixed the ring-polymer centroid on the new dividing surface. Configurations are sampled every 2 ps to serve as the initial position of the child trajectory used to calculate the flux-side correlation function. For each initial position, 100 individual trajectories are generated from different initial momenta sampled from the Boltzmann distribution. These trajectories then propagate unconstrained for 0.1 ps where the transmission coefficient reaches a plateau value.

## II.D Quantum dynamics

The time-dependent quantum wave packet method is used to calculate the thermal rate coefficients as well. The Hamiltonian of the system in the reactant Jacobi coordinates can be written as[53]

$$\hat{H} = -\frac{1}{2\mu_R}\frac{\partial^2}{\partial R^2} - \frac{1}{2\mu_r}\frac{\partial^2}{\partial r^2} + \frac{(\hat{J}_{tot} - \hat{j})^2}{2\mu_R R^2} + \frac{\hat{j}^2}{2\mu_r r^2} + \hat{V}(R, r, \theta), \tag{13}$$

where $\mu_R$ is the reduced mass between Cl and HCl; $\mu_r$ is the reduced mass of the reactant HCl; *R* is the distance from the attacking Cl atom to the center of mass of the reactant HCl; *r* is the bond distance of the reactant HCl; $\theta$ is the bending angle between the vectors ***R*** and ***r***; $\hat{J}_{tot}$ is the total angular momentum operator of the system; $\hat{j}$ is the rotational angular momentum operator of the reactant HCl; $\hat{V}$ is the potential energy operator.

The time-dependent wavefunction is expanded as

$$\Psi^{J_{tot}M\varepsilon} = \sum_{nvjK} C_{nvjK}^{J_{tot}M\varepsilon}(t) u_n^v(R) \phi_v(r) \Phi_{jK}^{J_{tot}M\varepsilon}(\hat{R}, \hat{r}), \tag{14}$$

where $C_{nvjK}^{J_{tot}M\varepsilon}(t)$ are time-dependent coefficients, *n* and *v* are labels for the basis functions along the coordinates *R* and *r*, respectively. *M* and *K* are projections of the total angular momentum *J* on the space-fixed (SF) and body-fixed (BF) *z* axes, respectively. The BF $z$ axis is defined to coincide with $R$. $\Phi_{jK}^{J_{tot}M\varepsilon}$ is the parity-adapted total angular momentum eigenfunction in the BF frame, which is defined as

$$\Phi_{jK}^{J_{tot}M\varepsilon} = \frac{1}{\sqrt{2(1+\delta_{K0})}}\sqrt{\frac{2J_{tot}+1}{8\pi}}\left[D_{KM}^{J_{tot}}P_{jK} + (-1)^{P} D_{KM}^{J_{tot}}P_{jK}\right], \tag{15}$$

where the total parity is $P=(-1)^{\varepsilon+J_{tot}}=+1$ in this calculation and $\varepsilon$ is the parity of system. $D_{KM}^{J_{tot}}$ is the Wigner rotation matrix[54]. $P_{jK}=\sqrt{2\pi}Y_j^K(\theta,0)$ are normalized associated Legendre polynomials.

The reaction probabilities for a specified initial state can be calculated at a dividing surface $r = r_s$,

$$P_{v_0J_0\tau}^{J_{tot}\varepsilon}(E_c) = \frac{\hbar}{\mu_r}\operatorname{Im}(\langle\psi_{iE}\,|\,\psi'_{iE}\rangle)\,|_{r=r_s}, \tag{16}$$

where $\psi_{iE}$ and $\psi'_{iE}$ are the time-independent wavefunction and its first derivative along $r$, which can be calculated by Fourier transforming the time-dependent wavefunction. ($v_0$ , $J_0$) and $\tau$ denote the initial rovibrational state and the parity of the reactant HCl. The integral cross section (ICS) from the initial state ($v_0$, $J_0$, $\tau$ ) is obtained by summing the reaction probabilities over all relevant partial waves

$$\sigma_{v_0J_0\tau}(E_c) = \frac{1}{2J_0+1}\frac{\pi}{2\mu E_c}\sum_{K_0\varepsilon}\sum_{J_{tot}\geq K_0}(2J_{tot}+1)P_{v_0J_0\tau K_0}^{J_{tot}\varepsilon}(E_c). \tag{17}$$

The initial state-specific rate constant is obtained by thermal averaging the collision energy of the corresponding ICS as

$$k_{v_0J_0\tau}(T) = \frac{1}{Q_e}\sqrt{\frac{8k_BT}{\pi\mu}}\frac{1}{(k_BT)^2}\int_0^\infty dE_cE_c\exp(-E_c/k_BT)\sigma_{v_0J_0\tau}(E_c), \tag{18}$$

where $E_c$ is the collision energy and $k_B$ is the Boltzmann constant. The electronic partition function $Q_e$ is given by $2+e^{-882.0/k_BT}$, in which the spin–orbit splitting of Cl is taken as 882.0 $cm^{-1}$.[5, 14, 46] The thermal rate constant is calculated by Boltzmann averaging of the initial state-specific rate constants as

$$k(T) = \frac{\sum_{v_0J_0}(2J_0+1)k_{v_0J_0\tau}(T)\exp(-E_{v_0J_0}/k_BT)}{\sum_{v_0J_0}(2J_0+1)\exp(-E_{v_0J_0}/k_BT)} \tag{19}$$

An L-shaped grid is used in this work.[55] The numerical parameters are listed in Table I. For the scattering coordinate $R$, 450 sine discrete variable representation (DVR) basis/points are used in the range from 1.5 to 25.0 $a_0$ and 210 sine DVR basis/points are used in the interaction region. For the dissociating $r$, 32 potential optimized DVR (PODVR) basis/points are used in the interaction region, and 8 PODVR basis/points are used in the asymptotic region. The initial vibrational states of the reactant HCl from $v_0 = 0$ - 3 are calculated, and for $v_0 = 0$, the initial rotational states from $J_0 = 0$ - 16 are calculated, for $v_0 = 1$ - 3, $J_0 = 0$ are calculated. The total angular momentum $J_{tot}$ is taken up to 350 to obtain converged ICSs. The angular basis is controlled by $j_{max} = 108$. The flux dividing surface is positioned at $r_s$ =2.65 Å. The total propagation time of wave packet is 55, 000 a.u. with a time step of 10.0 a.u.

# III. Results and discussion

## III.A Properties of the NN PES

As a compromise between efficiency and accuracy, the final PIPNN PES and EANN PES contains 20 and 20 neurons in the two hidden layers, resulting in 721 and 1024 fitted parameters, respectively. Figure 1 shows the diagram of the reaction path for Cl($^2$P) + HCl → ClH + Cl($^2$P) reaction, with geometries of the system at critical points. Along the reaction path, there is a van der Waals potential well between the asymptotic region and the TS. In this potential well, the Cl-H bond length is 1.28 Å, and the Cl atom is with 2.56 Å from the H atom of ClH molecule. The potential well depths on PIPNN PES and on EANN PES are 1.28 kcal/mol and 1.25 kcal/mol, respectively, matching properly with the MRCI value, 1.55 kcal/mol. In particular, the saddle point of the title reaction is found to be nonlinear, with the Angle of Cl-H-Cl as 136.9°, and the bond length of H-Cl bond as 1.499 Å. The state of the TS is $\tilde{X}^2B_1$, in $C_{2v}$ group, coincident with the finding of Schatz et al[8]. The barrier heights on PIPNN PES and EANN PES are 10.48 kcal/mol and 10.53 kcal/mol, respectively. Such nonlinear geometry of bottleneck is consistent with previously findings from D. Truhlar[1] and G. Schatz,[8] and with similar energetics. The heights of the barrier from the previous two PESs are basically the same as ours. The PIPNN PES fitting errors are shown in subplot of Figure 2, with small errors evenly distributed over the entire energy range. Its overall RMSE is 1.90 meV with the maximum deviation as 45.79 meV, an indication of excellent fitting performance. Figure 2 shows contour plots as functions of the breaking ($R_{HCl_1}$) and forming ($R_{HCl_2}$) bonds with the bond angle ∠ClHCl relaxed. It is clear to know that this is a typical symmetric reaction. TABLE I shows the comparison of energy, frequency and structure of each stable point of Cl($^2$P) + HCl → HCl + Cl($^2$P) reaction. In general, PIPNN PES and EANN PES results are highly consistent with MRCI-F12+Q/AVTZ level results due to minimal fitting errors. The TS structure of the three LEPS-type PES (BMCR PES, DIM-3C PES and PK3 PES) is collinear, which is inconsistent with the ab initio nonlinear TS structure, and the heights of the potential barrier are all too low, as about 8.50 kcal/mol. For POL-CI PES, the geometry for TS is correct, but still it has a low potential barrier of TS. The TS structure from DCBKS PES are close to the MRCI-F12+Q/AVTZ level result, but the barrier height with deviation about 25 %. Since the three semi-empirical potential energy surface TS structures are all linear, there are four vibration frequencies whose imaginary frequencies are not much different from *ab initio* results, but whose real frequencies differ by at least 40% or more. Due to the inaccuracy of the quantum chemical calculation method used by POL-CI PES, the real frequency of TS's vibration frequency is too high to 31%, while the virtual frequency is not much different. However, the real frequency difference of DCBKS PES is not much, but its virtual frequency difference is up to 10%.

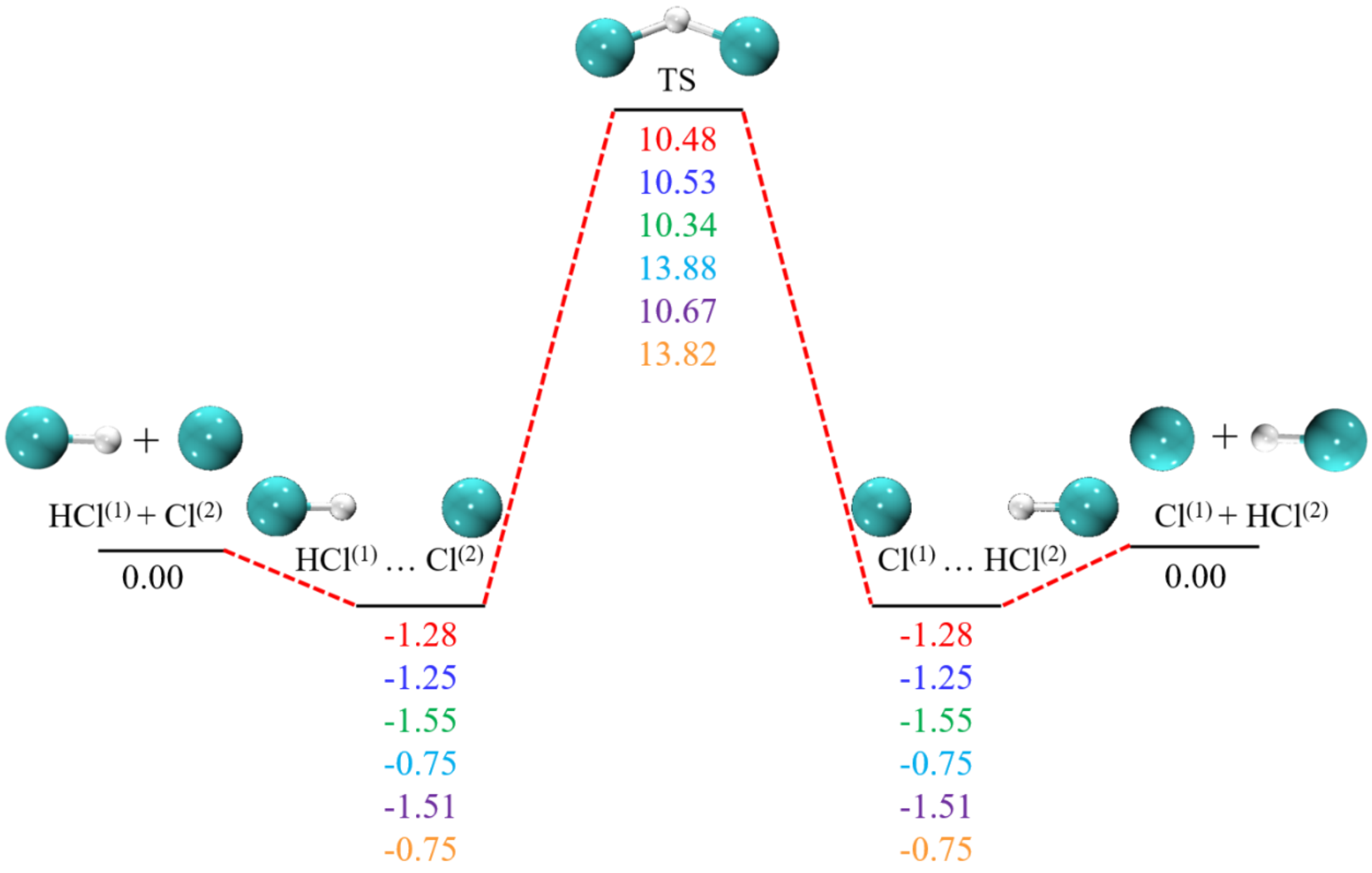

Figure 1 Schematic illustration of the title reaction energetics of the stationary points along the title reaction. All energies are in kcal mol$^{-1}$ and relative to the Cl($^2$P) + HCl reactant asymptote at various levels: PIP-NN, EANN, MRCI-F12+Q, MRCI-F12, MRCI+Q, MRCI, from top to bottom.

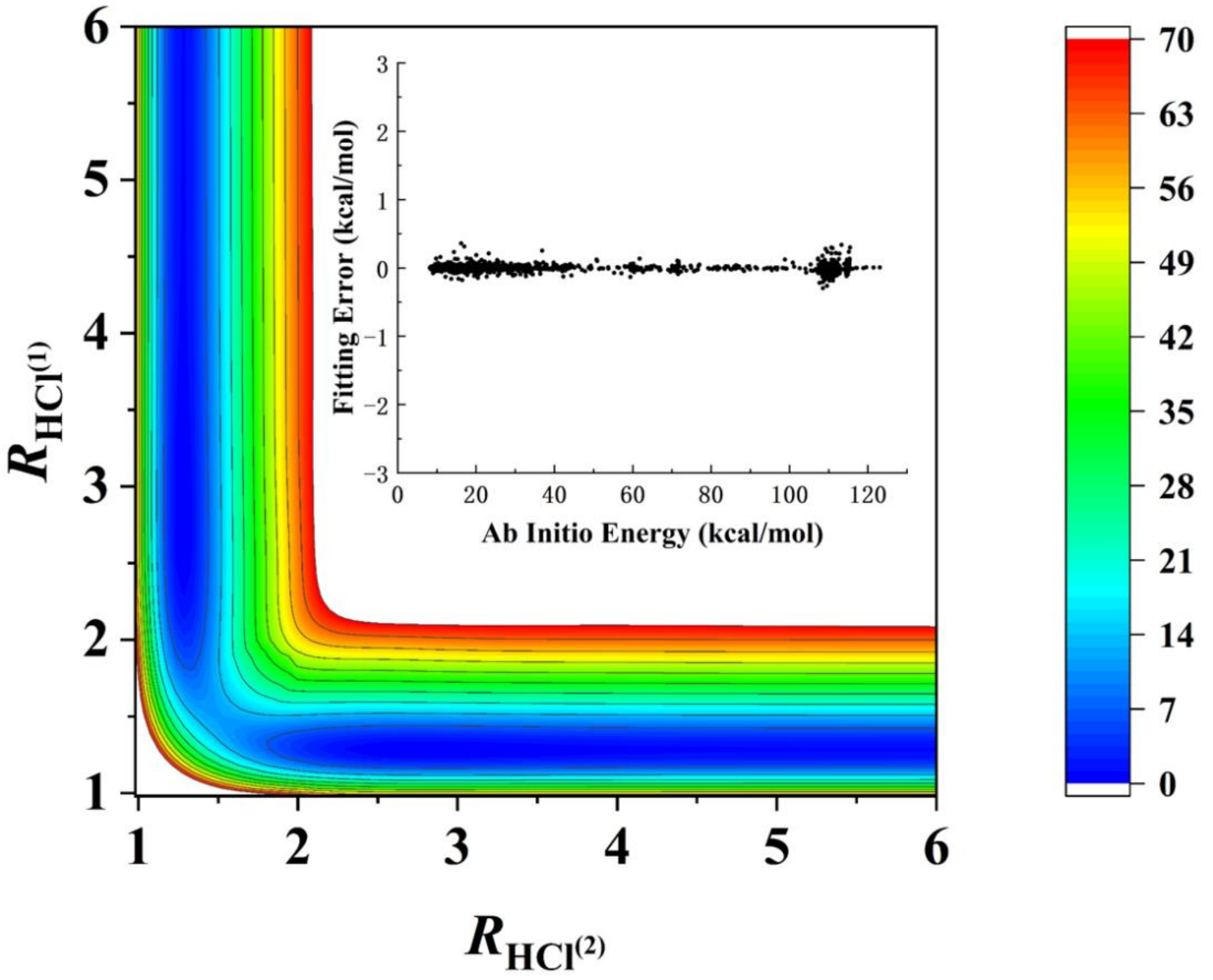

Figure 2 Contour plots as a function of the two reactive bonds $R_{HCl_1}$ and $R_{HCl_2}$ (bond in Å). The subplot shows the fitting errors (in kcal/mol) of the PIP-NN PES.

TABLE I Comparison of energy, frequency and structure at each stable point of Cl($^2$P) + HCl → HCl + Cl($^2$P) reaction.

| Species | Note | $R_{HCl}$ (Å) | <ClHCl (◦) | E (kcal/mol) | Frequency ($cm^{-1}$) | | | |
|---|---|---|---|---|---|---|---|---|
| | | | | | 1 | 2 | 3 | 4 |
| $Cl_1$ + $HCl_2$ | Ab initio[a] | 1.279 | | 0.00 | 2991 | | | |
| | Ab initio[b] | 1.279 | | 0.00 | 2981 | | | |
| | PIP-NN PES[c] | 1.279 | | 0.00 | 2991 | | | |
| | EANN PES[d] | 1.279 | | 0.00 | 2991 | | | |
| | BCMR[e] | 1.275 | | 0.00 | 2990 | | | |
| TS | Ab initio[a] | 1.499 | 136.9 | 10.34 | -1510 | 1113 | 204 | |
| | Ab initio[b] | 1.499 | 136.9 | 10.67 | -1506 | 1103 | 208 | |

| | | | | | | | |
|---|---|---|---|---|---|---|---|
| | PIP-NN PES[c] | 1.499 | 136.9 | 10.48 | -1520 | 1114 | 194 |
| | EANN PES[d] | 1.499 | 136.8 | 10.53 | -1381 | 1091 | 191 |
| | BCMR[e] | 1.467 | 180 | 8.55 | -1397 | 508/508 | 343 |
| | DIM-3C[f] | 1.488 | 180 | 8.30 | -1562 | 256/256 | 360 |
| | PK3[g] | 1.488 | 180 | 8.55 | -1467 | 691/691 | 345 |
| | POL-CI PES[h] | 1.473 | 161.4 | 6.34 | -1606 | 1617 | 326 |
| | DCBKS[i] | 1.498 | 137.7 | 9.87 | -1673 | 1099 | 206 |
| $HCl_1 + Cl_2$ | Ab initio[a] | 1.279 | | 0.00 | 2991 | | |
| | Ab initio[b] | 1.279 | | 0.00 | 2981 | | |
| | PIP-NN PES[c] | 1.279 | | 0.00 | 2991 | | |
| | EANN PES[d] | 1.279 | | 0.00 | 2991 | | |
| | BCMR[e] | 1.275 | | 0.00 | 2990 | | |

[a]This work,MRCI-F12+Q/AVTZ

[b]This work,MRCI+Q/AVTZ

[c]This work,PIP-NN PES

[d]This work, EANN PES

[e]The BCMR PES, see detail in Ref.[3]

[f]The DIM-3C PES, see detail in Ref.[6]

[g]The PK3 PES, see detail in Ref.[5]

[h]The POLCl PES, see detail in Ref.[1]

[i]The DCBKS PES, see detail in Ref.[8]

# III.B RPMD rate coefficients

In this work, the thermal rate coefficients for the Cl($^2$P) + XCl (X=H, D, Mu) reactions were calculated at the temperature range of 200-1000 K. This calculation is first performed with one bead, which provides the classical limit, and then the number of beads increases until converged.

In Cayley-RPMD calculations the number of beads needed depends on different temperatures and isotopes. The number of beads should not be less than the minimum suggested by Eq. (11). The rate coefficients calculated by the RPMD method and other theoretical methods as well as the rate coefficients measured experimentally are listed in Table II.

As described in the section II C, the reaction is in deep-tunneling region at 312.5 K, since the $T_c$ is 348 K for the title reaction. From previous discussions, the results below $T_c$ would underestimate the rate coefficients since the title reaction is with symmetric barrier.[56] But in this work, the Cayley-RPMD results are still in good agreement with experimental values.

The left panel of Figure 3 shows the PMF of Cl($^2$P) + XCl (X=H, D, Mu) reaction at 312.5 K. The convergent RPMD barriers (with the optimal number of beads) for all three reactions are lower than the classical (single-bead) results. This is due to tunneling effects that make it easier for the three isotopes to penetrate the potential energy barrier. The free energy barriers from different isotopic reactions decrease as $\Delta G_D > \Delta G_H > \Delta G_{Mu}$ according to the decrease in the mass of these isotopes. This order comes from the fact that the smaller the mass, the greater the tunneling capacity.

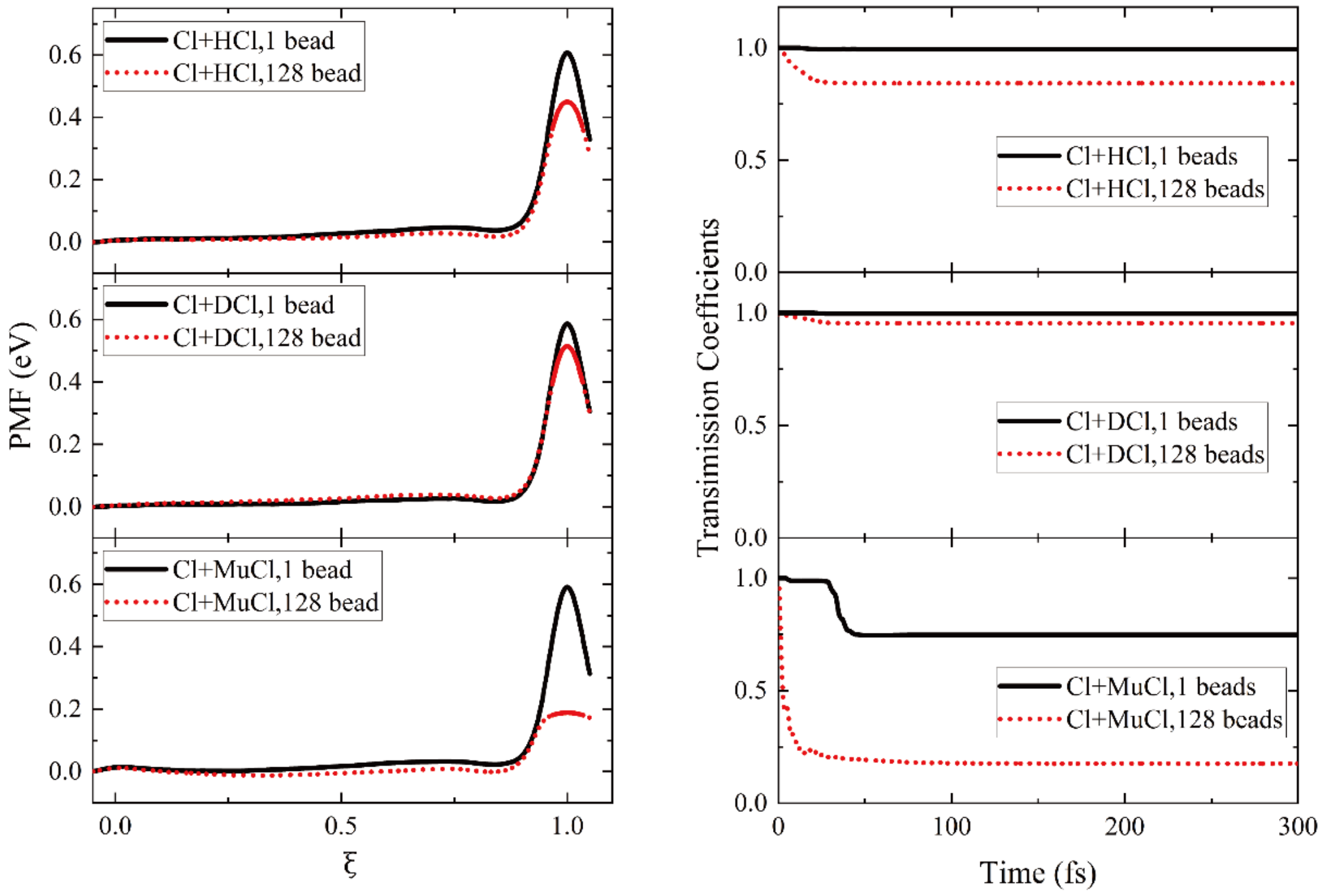

Figure 3 Potentials of mean force (PMF) (left panels) and transmission coefficients (right panels) of the Cl($^2$P) + XCl (X=H, D, Mu) reactions at 312.5 K.

The right panel of Figure 3 shows the corresponding transmission coefficient with time at 312.5 K. The curves of these three reactions converge rapidly within 20~30 fs, and the RPMD values are lower than their single-bead counterparts, indicating greater recrossing. The heavier the mass, the less tunneling and recrossing effect, indicating the larger transmission coefficient value. So the trend of $\kappa(t \to \infty)$ is $\kappa_{\mathrm{D}} > \kappa_{\mathrm{H}} > \kappa_{\mathrm{Mu}}$. Both single-bead and RPMD transmission coefficients of $Cl(^2P)$ + MuCl reaction have some oscillations before 50 fs, due to oscillation of the value of the reaction coordinate around the TS, as seen earlier for Mu + $H_2$ [57, 58] and H+$CH_4$[59] reactions, and without physical meaning. Table II shows that the converged RPMD transmission coefficients also increase with the increasing of temperature, indicating that the recrossing is more at low temperature, which is consistent with previous studies[39, 40]. As can be seen from the table II, the potential of mean force (PMF) of RPMD rates at temperatures 200 K to 1000 K, the free energy barrier increases with the increase in temperature, which is caused by the increase in kinetic energy of the system. The RPMD barrier heights for title reaction at 200, 312.5, 500, 1000 K are 0.364, 0.455, 0.561, and 0.720 kcal/mol, respectively. The transmission coefficient also increases with the increase of temperature, from 0.686 at 200 K to 0.867 at 1000 K, indicating less tunneling and recrossing with higher temperature.

TABLE II. Summary of RPMD rate coefficients of the $Cl(^2P)$ + XCl → XCl + $Cl(^2P)$ (X = H, D, Mu) reaction with other experimental and theoretical results. The rate coefficients are in $cm^3 \cdot molecules^{-1} \cdot s^{-1}$, ΔG is in eV, $\xi^{\neq}$ and $\kappa$ are dimensionless.

| $T$/K | 200 | 300 | 312.5 | 368.2 | 423.2 | 500 | 600 | 1000 |
|---|---|---|---|---|---|---|---|---|
| $n_{min}$ | 21.6 | 14.4 | 13.8 | 11.7 | 10.2 | 8.6 | 7.2 | 4.3 |
| Cl + HCl | | | | | | | | |
| $N_{\mathrm{beads}}$ | 128 | 128 | 128 | 64 | 64 | 64 | 16 | 16 |
| $\xi^{\neq}$ | 0.999 | 0.999 | 0.999 | 0.999 | 0.999 | 0.999 | 0.999 | 1.000 |
| $\Delta G(\xi^{\neq})$ | 0.364 | 0.448 | 0.455 | 0.502 | 0.538 | 0.561 | 0.586 | 0.720 |
| $k_{\mathrm{QTST}}$ | $2.62\times10^{-18}$ | $1.50\times10^{-16}$ | $2.46\times10^{-16}$ | $7.79\times10^{-16}$ | $2.42\times10^{-15}$ | $1.40\times10^{-14}$ | $8.46\times10^{-14}$ | $2.10\times10^{-12}$ |
| $\kappa$ | 0.686 | 0.839 | 0.846 | 0.880 | 0.893 | 0.901 | 0.897 | 0.867 |
| $k_{\mathrm{RPMD}}$[a] | $9.00\times10^{-19}$ | $6.25\times10^{-17}$ | $1.04\times10^{-16}$ | $3.37\times10^{-16}$ | $1.05\times10^{-15}$ | $6.08\times10^{-15}$ | $3.56\times10^{-14}$ | $8.01\times10^{-13}$ |
| $k_{\mathrm{RPMD}}$[b] | $2.67\times10^{-18}$ | $2.00\times10^{-16}$ | $3.05\times10^{-16}$ | $1.26\times10^{-15}$ | $3.37\times10^{-15}$ | … | $3.25\times10^{-14}$ | $3.97\times10^{-13}$ |
| $k_{\mathrm{ICVT/LCG2}}$[c] | $1.70\times10^{-17}$ | $7.50\times10^{-17}$ | $1.00\times10^{-15}$ | $3.30\times10^{-15}$ | $7.80\times10^{-15}$ | … | $4.70\times10^{-14}$ | $3.00\times10^{-13}$ |
| $k_{expt}$[d] | … | … | $(1.5\pm0.8)\times10^{-15}$ | $(5.1\pm2.4)\times10^{-15}$ | $(1.5\pm0.6)\times10^{-14}$ | $3.89\times10^{-14}$ | … | … |
| Cl + DCl | | | | | | | | |
| $N_{\mathrm{beads}}$ | 128 | 128 | 128 | 64 | 64 | 64 | 16 | 16 |
| $\xi^{\neq}$ | 0.999 | 0.999 | 0.999 | 0.999 | 0.999 | 0.999 | 0.999 | 1.000 |
| $\Delta G(\xi^{\neq})$ | 0.428 | 0.506 | 0.509 | 0.553 | 0.583 | 0.600 | 0.633 | 0.784 |
| $k_{\mathrm{QTST}}$ | $6.72\times10^{-20}$ | $1.47\times10^{-17}$ | $3.00\times10^{-17}$ | $1.41\times10^{-16}$ | $6.92\times10^{-15}$ | $5.51\times10^{-15}$ | $3.28\times10^{-14}$ | $9.93\times10^{-13}$ |
| $\kappa$ | 0.881 | 0.963 | 0.964 | 0.969 | 0.970 | 0.967 | 0.951 | 0.915 |
| $k_{\mathrm{RPMD}}$[a] | $2.96\times10^{-20}$ | $7.03\times10^{-18}$ | $1.44\times10^{-17}$ | $6.73\times10^{-17}$ | $3.27\times10^{-16}$ | $2.56\times10^{-15}$ | $1.46\times10^{-14}$ | $3.98\times10^{-13}$ |
| $k_{\mathrm{RPMD}}$[b] | $1.97\times10^{-19}$ | $3.76\times10^{-17}$ | $5.60\times10^{-17}$ | $2.88\times10^{-16}$ | $1.00\times10^{-15}$ | … | $1.42\times10^{-14}$ | $2.37\times10^{-13}$ |
| $k_{\mathrm{ICVT/LCG2}}$[c] | $1.60\times10^{-18}$ | $1.60\times10^{-16}$ | $2.40\times10^{-16}$ | $9.80\times10^{-16}$ | $2.80\times10^{-15}$ | … | $2.30\times10^{-14}$ | $2.10\times10^{-13}$ |
| $k_{expt}$[d] | … | … | $(1.7\pm0.8)\times10^{-16}$ | $(1.0\pm0.5)\times10^{-15}$ | $(3.6\pm1.6)\times10^{-15}$ | $1.37\times10^{-14}$ | … | … |
| Cl+MuCl | | | | | | | | |
| $N_{\mathrm{beads}}$ | 128 | … | 128 | … | … | … | 16 | 16 |
| $\xi^{\neq}$ | 1.000 | … | 1.000 | … | … | … | 0.999 | 0.999 |
| $\Delta G(\xi^{\neq})$ | 0.130 | … | 0.188 | … | … | … | 0.318 | 0.480 |
| $k_{\mathrm{QTST}}$ | $2.86\times10^{-12}$ | … | $6.21\times10^{-12}$ | … | … | … | $1.89\times10^{-11}$ | $4.52\times10^{-11}$ |
| $\kappa$ | 0.167 | … | 0.176 | … | … | … | 0.240 | 0.246 |
| $k_{\mathrm{RPMD}}$[a] | $2.39\times10^{-13}$ | … | $5.41\times10^{-13}$ | … | … | … | $2.13\times10^{-12}$ | $4.88\times10^{-12}$ |
| $k_{\mathrm{RPMD}}$[b] | $4.05\times10^{-14}$ | … | $2.94\times10^{-13}$ | … | … | … | $1.69\times10^{-12}$ | $5.13\times10^{-12}$ |

[a]This work using PIP-NN PES

[b]Reference 16 using BCMR PES[3]

[c]Reference 1 using POLCl PES[1]

[d]Reference 1[1]

The left panel of Figure 4 shows Arrhenius plots comparing Cayley-RPMD rate coefficients of Cl($^2$P) + HCl reactions with experimental values and other theoretical results. The ICVT rate coefficients are with quantum correction by *Garrett et al*[1]. The deviation between the results from ICVT is large, denoted as ICVT/LCG1 and ICVT/LCG2, although both from POL-CI PES, the latter scales the potential with a parameter 1.42 to match the experimental value. Previous standard RPMD results from our group[14] based on BCMR PES were smaller than the experimental values. The experimental results are collected in the work of Klein *et al*[60] and Kneba and Wolfrum[61], denoted as Expt. The Cayley-RPMD results on PIP-NN PES and EANN PES are still smaller than the experimental ones. However, when we scaled the PIP-NN PES with 0.815, the result of Cayley-RPMD was the closest to the experimental value, with deviation less than 20%.

Quantum wave packet calculations could provide accurate thermal rate constants in condition that all relevant rovibrational states are thermally averaged. In the temperature range from 200 K to 1000 K, the reactant rotational excitations up to $J_{\mathrm{tot}} = 16$ have significant contributions to the thermal rate constants. By contrast, vibrational excitations just have small contributions at high temperatures above 500 K. Therefore, the states with simultaneous vibration-rotation excitation are not considered in the quantum wave packet calculations, which is believed to have negligible effect on the thermal rate constants. As shown in Figure 4, the quantum thermal rate constants agree reasonably well with the Cayley-RPMD results although the former are slightly higher at low temperatures. QD calculations show that, on one hand, the reaction energy threshold is visibly lower than the classical barrier height, indicating the existence of significant quantum tunneling effect for the reactions, which consistent with the PMF curves from RPMD, as shown in Figure 3, the left panel. On the other hand, the low-lying rotationally excited states of HCl with $J_0 \leq 4$ inhibit the reaction. As the rotational quantum numbers increases, the rotational excitation starts to promote the reaction. In addition, some sharp peaks emerge in the vibrationally exited reaction probability curves, as can be seen in the Figure S1 in Supporting Information. These peaks are very likely to be caused by dynamical resonances. As is well-known, dynamical resonances generally induce quantum tunneling, causing the enlargement of rate constants. Overall, RPMD does not include resonance effect, and approximates quantum effect. In contrast, QD calculations don't include all thermal internal states due to large computational efforts. These approximations possibly bring the slight difference between QD and Cayley-RPMD rate constants at low temperatures. And for the experimental results, it has known that they overestimate the rate coefficients, due to both the side reactions such as $Cl+H_2$ and the finite pressure effect. The approximations in theoretical calculations and other effects in experimental measurement all possibly create uncertainty on the results.

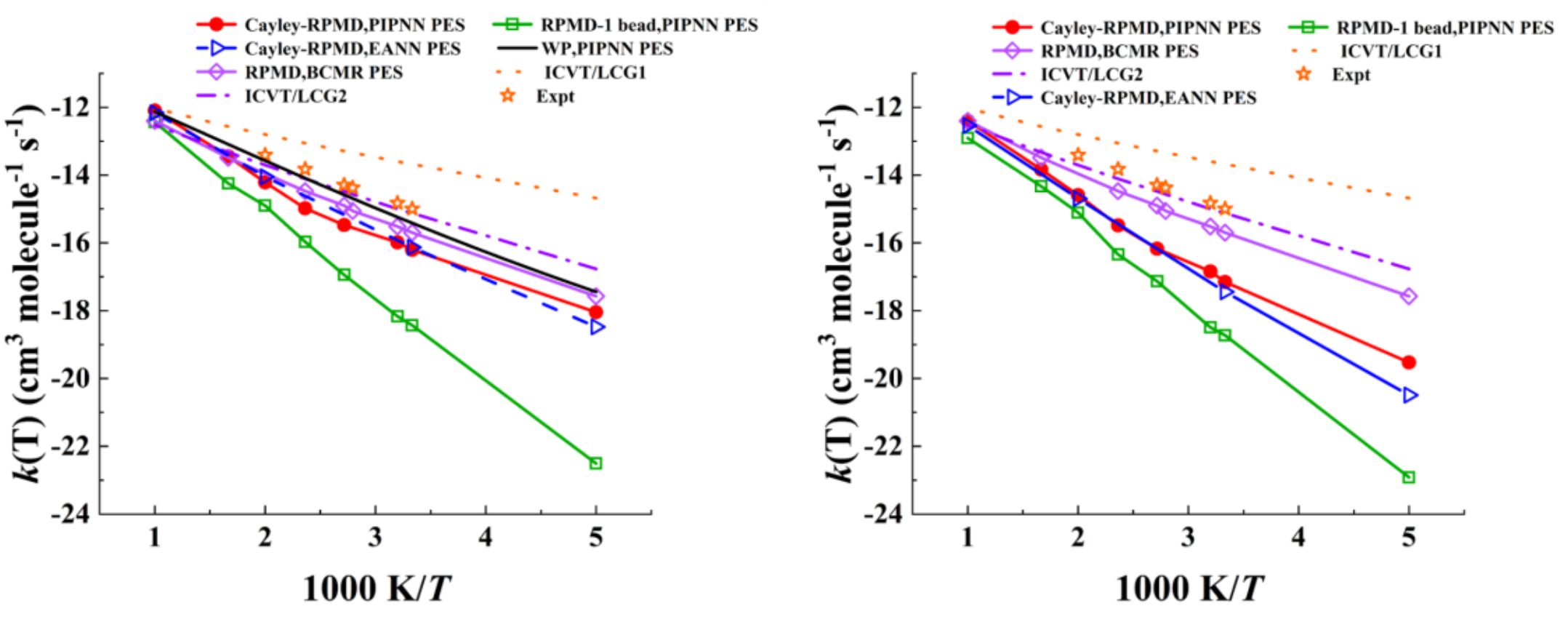

Figure 4 Left panel: Comparison of rate coefficients for the Cl($^2$P) + HCl reaction obtained with Cayley-RPMD on PIP-NN and EANN PESs, standard RPMD on BCMR PES[14], quantum dynamics using wave packet based on PIP-NNPES (denoted as WP), ICVT method[1] (on POL-CI PES denoted as ICVT/LCG1, scaled POL-CI PES denoted as ICVT/LCG2), and experiments[60, 61]. Right panel: Comparison of rate coefficients for the Cl($^2$P) + DCl reaction obtained with Cayley-RPMD on PIP-NN and EANN PESs, standard RPMD on BCMR PES[14], ICVT method[1](Based on POL-CI PES denoted as ICVT/LCG1,scaled POL-CI PES denoted as ICVT/LCG2), and experiments[60, 61].

The right panels of Figure 4 shows Cl($^2$P) + DCl rate coefficients in the Arrhenius plot. All theoretical values are smaller than experimental values, except for the result of ICVT/LCG1. The ICVT results of the two groups are significantly different again, with the result of ICVT/LCG2 still closer to the experimental value than the result of ICVT/LCG1. The RPMD results are still smaller than experimental values. The deviation would come from the same sources discussed above, and the unexpected match between ICVT/LCG2 results and the experimental ones are quite likely from the cancellation between errors, since that work was on a PES containing colinear TS and smaller potential barrier, and the method used is based on TST, with only empirical correction of quantum effects.

It's noteworthy that, the RPMD results from PIP-NN PES and EANN PES deviate visibly from each other at lower temperatures of 200 and 300 K.This may stem from the difference between the shape of both PESs, as reported in Cl+$CH_4$[62]. From Table 1, one can find that the imaginary and the largest frequencies at TS on the two PESs are different. The absolute values of those frequencies from the PIP-NN PES are slightly larger than MRCI values, whereas those from the EANN PES are smaller. In addition, deuterium has less tunneling effect than hydrogen at lower temperature, which would enlarge the difference of the results on the two PESs.

For the Cl($^2$P) + MuCl reaction, the Cayley-RPMD results on PIP-NN PES and standard RPMD on BCMR PES[14] are only listed in Table II since there is no experimental result till now. both sets of results are consistent at high temperatures, but deviate at low temperatures. And at lower temperatures, the Cayley-RPMD results on PIP-NN PES are obviously larger than those from BCMR PES.

The KIE of H and D isotopes is shown in Figure 5. The experimental values are collected from Garrett *et al*[1], and other theoretical values are taken from work by Bondi *et al.*[3] The calculated KIE values converge to two at temperature as high as 1000 K, but diverge at low temperatures. The Cayley-RPMD results on PIP-NN PES are highly consistent with the experimental ones over the whole temperature range with relative error within 22%. This can be understood as the RPMD can take into account the ZPE accurately, which is usually significantly affect the KIE, as discussed in our previous work[63]. Furthermore, the PIP-NN PES performs better on describing both the energies and frequencies of stationary points.

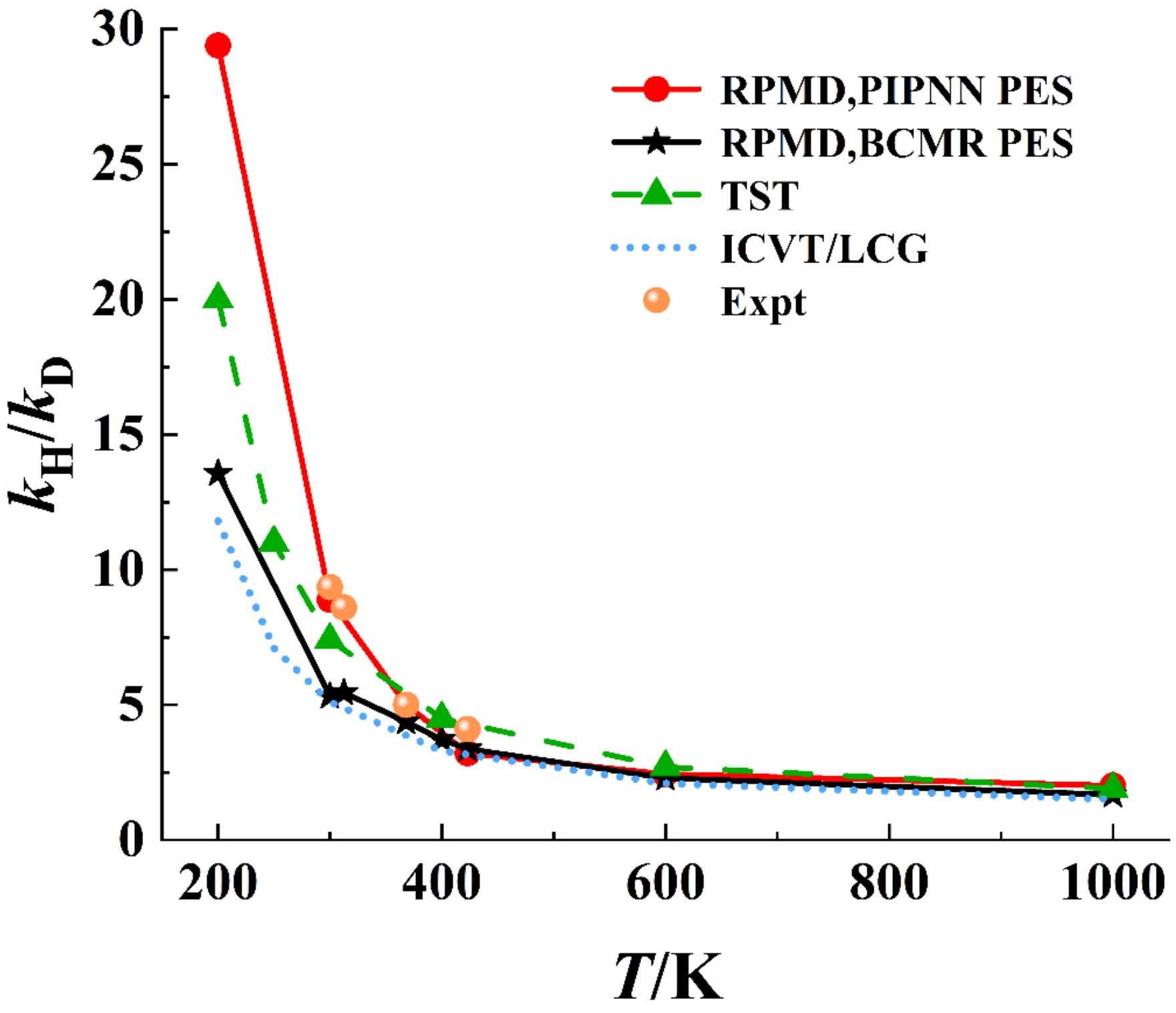

Figure 5 Comparison between calculated KIEs ($k_H/k_D$) and measured ones. Comparison of $k_H/k_D$ among Cayley-RPMD (PIPNN PES), PRMD (BCMR PES)[14], TST[1], ICVT/LCG2[1] and experiments[60, 61].

## IV. Summary and conclusions

In this work, we have investigate the reaction dynamics of Cl($^2$P) + HCl → HCl + Cl($^2$P). Firstly, we have developed two new full-dimensional neural network PESs for ground state of the title reaction, the permutation invariant polynomial neural network (PIP-NN) method and embedded atomic neural network (EANN) method, based on 5986 points and 6515 points at MRCI-F12+Q/AVTZ level, and with the total root mean square error was 0.043 kcal/mol and 0.056 kcal/mol, respectively. In particular, the EANN method is the first application of gas phase bimolecular reactions. From the comparison, the performance is well, and is promising for being applied into multiple atomic reactions with active learning technique. We have also confirmed the TS of the title reaction is nonlinear. Then the full-dimensional approximate quantum mechanical method, ring-polymer molecular dynamics with Cayley propagator (Cayley-RPMD), is used for obtaining thermal rate coefficients for the isotopic reactions Cl($^2$P)+XCl, (X=H, D, Mu) within 200 K-1000 K. Especially, for the hydrogen isotopic reaction, we also performed QD with wave packet to obtain the thermal rate coefficients for comparison. For all the three isotopic reactions, the PMF curves and transmission coefficients show that the title reaction is affected by recrossing and quantum mechanical effects, such as tunneling and ZPE. Comparing with other data from theoretical calculation and experiments, we find that except for results from ICVT/LCG1, all theoretical results are smaller than experimental measurements. Our results from both RPMD and QD can reproduce the experimental values reasonably well, and the results from RPMD are systematically lower than those from QD, due to different treatment of thermal average. The KIE results from our work are in the best agreement with the experiment comparing with other theoretical results, showing the perfect including of ZPE effect in RPMD. The deviation between rate coefficients from theoretical calculation and from experiments is attributed to the side reactions, finite pressure effect, and the activation by light and container. But since there is lack of more determining experimental results, this deviation cannot be definitely clarified. We hope that in near future, there can be more accurate experimental results for title reaction to verify our results.

## SUPPLEMENTARY MATERIAL

See supplementary material for details of table of parameters used in the wave packet calculations; Plots of reaction probabilities from different vibrational states of the reactant HCl; Plot of convergence of PMF curves with the changing of the number of beads at 312.5 K; Plots of the PMF curves, Transmission coefficients and Rate coefficients from Cayley-RPMD on PIP-NN PES with different $\Delta t$ at 312.5 K; Table of the PMF, transmission coefficient and rate coefficient from Cayley-RPMD on PIP-NN PES with different $\Delta t$ at 312.5 K, and percentage relative errors with respect to 0.1fs; Table of comparison of energy, frequency and structural parameters at each stable point of Cl ($^2$P) + DCl → DCl + Cl($^2$P) reaction.

## ACKNOWLEDGMENTS

This work is supported by the National Nature Science Foundation of China (No. 21973109 and 22173057), the research grant (No. 21JC1402700) from Science and Technology Commission of Shanghai Municipality, and Key Research Project of Zhejiang Laboratory (No. 2021PE0AC02). We also gratefully acknowledge

HZWTECH for providing computation facilities. The authors also thank helpful discussions from Wenbin Fan in Fudan University, Changjian Xie in Northwest University, Xixi Hu in Nanjing University and Jun Li in Chongqing University.

## Author Contributions

Qiang Li and Mingjuan Yang contributed equally to this work.

## DATA AVAILABILITY

The data that support the findings of this study are available from the corresponding author upon reasonable request.